\documentclass[aps,prb,showpacs,twocolumn,floats,psfig]{revtex4}
\usepackage{amssymb}
\usepackage{amsbsy}
\usepackage{amsmath}
\usepackage{epsfig}
\newcommand\bea{\begin{eqnarray}}
\newcommand\eea{\end{eqnarray}}
\newcommand\beq{\begin{equation}}
\newcommand\eeq{\end{equation}}

\newcommand{\non}{\nonumber}
\newcommand{\al}{\alpha}

\newcommand{\ga}{\gamma}

\newcommand{\si}{\sigma}

\newcommand{\la}{\langle}
\newcommand{\ra}{\rangle}

\newcommand{\f}{\frac}
\newcommand{\lam}{\lambda}

\begin{document}

\title {Entanglement production due to quench dynamics of an anisotropic $XY$
chain in a transverse field}

\author{K. Sengupta$^{1,2}$ and Diptiman Sen$^3$}

\affiliation{$^1$Theoretical Physics Division, Indian Association for the
Cultivation of Sciences, Jadavpur, Kolkata 700 032, India \\
$^2$TCMP division, Saha Institute of Nuclear Physics, 1/AF Bidhannagar,
Kolkata 700 064, India \\
$^3$Center for High Energy Physics, Indian Institute of Science, Bangalore
560 012, India}

\date{\today}

\begin{abstract}
We compute concurrence and negativity as measures of two-spin entanglement
generated by a power-law quench (characterized by a rate $\tau^{-1}$ and an
exponent $\al$) which takes an anisotropic $XY$ chain in a transverse field
through a quantum critical point (QCP). We show that only spins separated by
an even number of lattice spacings get entangled in such a process. Moreover,
there is a critical rate of quench, $\tau_c^{-1}$, above which no two-spin
entanglement is generated; the entire entanglement is multipartite. The ratio
of the entanglements between consecutive even neighbors can be tuned by
changing the quench rate. We also show that for large $\tau$, the concurrence
(negativity) scales as $\sqrt{\al/\tau}$ ($\al/\tau$), and we relate this
scaling behavior to defect production by the quench through a QCP.
\end{abstract}


\maketitle

\section {Introduction}

The role of entanglement in the theory of quantum phase transitions
has been a subject of recent studies [\onlinecite{osterloh}]. A number of
such works have pointed out that entanglement can be used as a tool
to characterize quantum phase transitions for both clean and
disordered systems [\onlinecite{entang1,rev1,rev2,as1,ib1}]. These works
computed either single-site (or single block) von Neumann entropies
or two-spin entanglement measures such as concurrence [\onlinecite{concref}]
and negativity [\onlinecite{negref}], and demonstrated that these measures
or their derivatives display a peak (discontinuous jump) either near
or at the second order (first order) quantum critical point (QCP).
They can therefore serve as tools for identifying quantum phase
transitions in equilibrium quantum critical systems. These studies
provide a bridge between quantum information theory and
equilibrium quantum critical phenomena which can be useful for
several aspects of quantum computations, cryptography and
teleportation [\onlinecite{qcomp}]. However, in all of these studies, the
computed entanglement receives contribution from the ground state of
the systems alone; excited states are not probed.

Recently, enormous progress has been made in understanding defect production 
due to non-equilibrium dynamics of a system passing through a critical point 
[\onlinecite{kibble1,antu,zur,an1}]. In particular, it was shown
that for a slow linear quench through a QCP, the defect density $n$
scales with the quench time $\tau$ with an universal exponent: $n \sim
\tau^{-\nu d/(z \nu+1)}$, where $\nu$ and $z$ are the correlation
length and dynamical critical exponents associated with the phase
transition and $d$ is the system dimension [\onlinecite{antu,an1}]. Such 
results have been extended to cases where the quench takes the system
through a quantum critical surface [\onlinecite{ds1}], along a gapless line
or a multicritical point [\onlinecite{ds2}], and for non-linear power-law 
quenches [\onlinecite{ds3}]. More recently, the two-spin entanglement 
properties of a quantum
system during time evolution after a sudden quench through a critical point
have been studied [\onlinecite{as2}]. However, the nature of two-spin
entanglement generation due to a finite rate of quench through a
quantum critical point has not been investigated so far, although other
kinds of entanglement have been studied [\onlinecite{cincio}].

In this work, we compute concurrence and negativity as measures of
two-spin entanglement generated by a power-law quench, characterized
by a rate $\tau$ and an exponent $\al$, which takes a spin-1/2 $XY$
chain in a transverse field through a QCP. Our central results are
as follows. First, we show that, in contrast to the studies of
entanglement in anisotropic $XY$ chains so far [\onlinecite{entang1}], such
a quench generates entanglement only between spins separated by an
even number of lattice spacings (even neighbors); the nearest
neighbor sites are not entangled. Second, a critical quench rate
$\tau_c^{-1}$ is required to generate two-spin entanglement (unlike
other kinds of entanglement which do not seem to require a critical
quench rate [\onlinecite{cincio}]). For faster quench rates ($\tau <
\tau_c$), there is no entanglement between any pair of sites, and
the entire entanglement is multipartite which is rather unusual.
Third, by tuning the quench rate, one can control the amount of
entanglement, and tune the ratio of entanglements between a spin and
its consecutive even neighbors to take values between zero and 1.
Finally, for large quench time $\tau$ and a given power $\al$, the
concurrence (negativity) scales as $\sqrt{\al/\tau}$ ($\al/\tau$);
this scaling is directly related to defect production by the quench
through the QCP. To the best of our knowledge, the scaling behavior
of the two-spin entanglement generated by a slow quench through a
QCP and its selective generation by tuning the quench rate have not
been reported so far. Hence our study constitutes a significant
extension of our current understanding of the nature of entanglement
in many-body systems. This study may be relevant for the generation
of entanglement as a resource for many algorithms of quantum
computations, cryptography and teleportation [\onlinecite{qcomp}].

The organization of the rest of this paper is as follows. In Sec. 
\ref{entangsec1}, we compute concurrence and negativity as measures of 
entanglement for the $XY$ model. This is followed by a discussion of our 
main results in Sec. \ref{results1}. We end with some concluding 
remarks in Sec. \ref{conclu1}.

\section{Measures of entanglement}
\label{entangsec1}

We begin with the Hamiltonian of the spin-1/2 anisotropic $XY$ spin
chain given by \beq H = \frac{J}{4} \sum_n [(1+\ga) \si^x_n
\si^x_{n+1} + (1-\ga)\si^y_n \si^y_{n+1} + h \si^z_n], \label{h1}
\eeq where $\si^a$ for $a=x,y,z$ are the Pauli matrices,
$J(1+(-)\ga)/2$ are interaction strengths between $x(y)$ components
of the nearest-neighbor spins (we set the lattice spacing $d=1$),
$h$ is the magnetic field in units of $J$ [we henceforth set
$\hbar=J=1$, and measure all energies (times) in units of $J
~(\hbar/J)$], and $\ga$ is the anisotropy parameter which varies
between $0$ (isotropic $XY$ chain) and $1$ (Ising limit). The phase
diagram of the above model can easily be obtained by mapping the
spins to fermions via a Jordan-Wigner transformation: $\si_i^z=
n_i-1/2$, and $c_i(c_i^{\dagger}) = \prod_{j=-\infty}^{i-1} \si_j^z
(-1)^i \si_i^{-} (\si_i^+)$ [\onlinecite{ds4,lev1}]. The fermions can be
shown to have an effective two-level Hamiltonian for each pair of
momenta $\pm k$ in terms of the states $|0 \rangle$ and
$|k,-k\rangle = c_k^\dagger c_{-k}^\dagger |0\rangle$ given by $H_k
= -~[ h ~+~ 2 \cos k] ~(I +\tau_z) - 2 \tau_y \ga \sin k$
[\onlinecite{ds4,lev1}], where $I$, $\tau_z$ and $\tau_y$ denote identity
and Pauli matrices in the $|0 \rangle$, $|k,-k\rangle$ space. The
equilibrium phase diagram for the model is well-known; for $|h| >
1$, there is a paramagnetic phase with $\la \si_i^x \ra = 0$, while
for $|h| < 1$, one finds a ferromagnetic phase with $\la \si_i^x \ra
\ne 0$. At $h= \pm 1$, there is a second order quantum phase
transition with $z=\nu=1$. At the transition at $h=1(-1)$, the
fermionic modes are gapless for $k=0(\pi)$. The quench dynamics of
the model, for a power-law time variation of the magnetic field
$h(t) = h |t/\tau|^{\al} {\rm sgn}(t)$, where ${\rm sgn}$ is the
signum function, has also been studied [\onlinecite{lev1,ds3}]. The quench
starts with all spins down at $t \to - \infty$ (i.e., the state
$|0\rangle$ for all $k$) and ends at $t \to \infty$ with a state in
which the probabilities of $|0\rangle$ and $|k,-k\rangle$ are given
by $p_k$ and $1 - p_k$ respectively. Here $p_k$ is the defect
formation probability and is given by $p_k= \exp(-\pi \tau_{\rm eff}
\ga^2 \sin^2 k)$, where $\tau_{\rm eff}= \tau/\al$ [\onlinecite{lev1,ds3}].

Armed with these results, we now compute the concurrence and
negativity as measures of the two-spin entanglement of the spin chain
generated by the quench. We note at the outset that the ground
states of the initial and final Hamiltonians, at the beginning and
end of such a quench process, are paramagnetic and do not posses any
two-spin entanglement. Thus we expect that both for very fast ($\tau
\to 0$) and very slow ($\tau \to \infty$) quenches, where the system
retains information only about the initial and final ground states,
the two-spin entanglement will vanish. Hence any finite entanglement
obtained after such a quench with a finite rate $\tau$ must be generated
by the non-adiabatic quench process and must therefore have contributions
from excited states of the system. To compute the concurrence and negativity,
we first note that the two-spin density matrix of the spin chain for any
two sites $i$ and $j=i+n$ is given by [\onlinecite{syl1}]
\bea \rho^{n} ~=~ \left( \begin{array}{cccc}
a_+^{n} & 0 & 0 & b_1^n \\
0 & a_0^n & b_2^n & 0 \\
0 & b_2^{n*} & a_0^n & 0 \\
b_1^{n*} & 0 & 0 & a_-^n \end{array} \right), \label{rhoij} \eea
where the matrix elements $a_{\pm}^n$, $a_0^n$ and $b_{1,2}^n$ can
be expressed in terms of the two-spin correlation functions 
\bea a_{\pm}^n &=& \la \f{1}{4} (1\pm \si_i^z) (1\pm \si_{i+n}^z) \ra , \non \\
a_0^n &=& \la \f{1}{4} (1 \pm \si_i^z) (1 \mp \si_{i+n}^z) \ra, \non \\ 
b_{1(2)}^n &=& \la \si_i^- \si_{i+n}^{-(+)} \ra \label{corr1}. \eea 
The symmetry under $\si^x_i \to - \si^x_i, ~\si^y_i \to - \si^y_n, ~ 
\si^z_i \to \si^z_i$ ensures that all correlation functions such as 
$\la \si^{\pm}_i \si_{i+n}^z\ra$ and hence the remaining matrix elements 
are zero. The non-zero correlation functions for an arbitrary non-linear 
quench can be computed by generalizing the method developed for a linear 
quench in Ref. [\onlinecite{lev1}]. We define 
\bea \al_n &=& \int_0^{\pi} \frac{dk}{\pi} p_{k} \cos(nk), \eea 
and note that since $p_k$ is invariant under $k \to \pi -k$, $\al_n=0$ when 
$n$ is odd. In terms of $\al_n$, the diagonal correlation functions are given 
by 
\bea \la \si_i^z \ra &=& 1 ~-~ 2 \al_0 , \non \\ 
\la \si_i^z \si_{i+n}^z \ra &=& \la \si_i^z \ra^2 - 4 \al_n^2 . 
\label{corr2} \eea 
Thus, for any two spins separated by odd number of lattice spacings, $\la 
\si_i^z \si_{i+n}^z \ra=\la \si_i^z \ra^2$ [\onlinecite{lev1}]. The 
off-diagonal 
correlators $\la \si_i^{a} \si_{i+n}^{b}\ra$ (where $a$, $b$ can take the 
values $+$, $-$) can also be computed in terms of $\al_n$. We shall present
explicit expressions for these for $n \le 6$ and provide a
qualitative discussion for large $n$ later. We find that $\la
\si_i^{\pm} \si_{i+n}^{\pm}\ra =b_1^n = 0$ for all $n$ since these
involve correlations between two fermionic annihilation or creation
operators and hence vanish. Further, $\la \si_i^{\pm}
\si_{i+n}^{\mp}\ra=b_2^n=0$ for all odd $n$ since these are odd
under the transformation $\si^x_n \to (-1)^n \si^x_n, ~\si^y_n \to
(-1)^n \si^y_n, ~\si^z_n \to \si^z_n$ which changes $J_{x,y} \to -
J_{x,y}$ and leaves $p_k$ invariant. For even $n\le 6$, we find
\bea \la \si_i^+ \si_{i+2}^- \ra &=& \al_2 \la \si_i^z \ra, \non \\
\la\si_i^+ \si_{i+4}^- \ra &=& (\al_4 \la \si_i^z \ra - 2 \al_2^2)
\la \si_i^z \si_{i+2}^z \ra, \non \\
\la \si_i^+ \si_{i+6}^- \ra &=& \Big[ \al_6\la \si_i^z
\si_{i+2}^z\ra +4\al_2(\al_2^2 +\al_4^2 -\al_4 \la
\si_i^z \ra) \Big] \times \non \\
&& \Big[ \la \si_i^z \ra[\la \si_i^z \si_{i+2}^z\ra -4 (\al_2^2 +
\al_4^2)] +16 \al_2^2 \al_4 \Big]. \non \\ \label{corr3} \eea 
Using these correlation functions, we can find all the non-zero matrix
elements of $\rho^n$ for $n\le 6$ and hence compute the concurrence
and negativity. The concurrence is given by ${\cal C}^n = {\rm max}
\{ 0, \sqrt{\lam_1^n} - \sqrt{\lam_2^n} - \sqrt{\lam_3^n} -
\sqrt{\lam_4^n} \}$, where $\lam_i^n$'s are the eigenvalues of
$\rho^n (\si^y \otimes \si^y \rho^{n*} \si^y \otimes \si^y) $ in
decreasing order [\onlinecite{concref}]. The $\sqrt{\lam_i^n}$ are given by
$\sqrt{a_+^n a_-^n}$ (appearing twice), and $a_0^n \pm |b_2^n|$.
Thus the spin chain has a non-zero concurrence if $|b_2^n| >
\sqrt{a_+^n a_-^n}$ given by \bea {\cal C}^n &=& {\rm
max}\left\{0,2(|b_2^n|-\sqrt{a_+^n a_-^n})\right\} \label{conc1}
\eea To compute the negativity ${\cal N}^n$, we need to take a
partial transpose of $\rho^n$ with respect to the labels
corresponding to the site $j=i+n$ in Eq. (\ref{rhoij})
[\onlinecite{negref}]. This interchanges $b_1^n \leftrightarrows b_2^n$; the
eigenvalues of the resultant matrix ${\bar \rho}^{n}$ are given by
${\tilde \lambda}_0^n= a_0^n$ (appearing twice), and ${\tilde
\lambda}_{\pm}^n=(1/2) [ a_+^n + a_-^n \pm \sqrt{(a_+^n - a_-^n)^2 +
4 |b_2^n|^2}]$ of which only ${\tilde \lambda}_{-}^n$ can become
negative. This happens when $ |b_2^n| > \sqrt{a_+^n a_-^n}$ and
yields \bea {\cal N}^n &=& {\rm max}\left\{0,|{\tilde
\lambda}_-^n|\right\}. \label{nega1} \eea In the next section, we
shall discuss the implications of these measures of entanglement in
the context of $XY$ model in a transverse field.

\begin{figure} \rotatebox{0}{\includegraphics*[width=\linewidth]{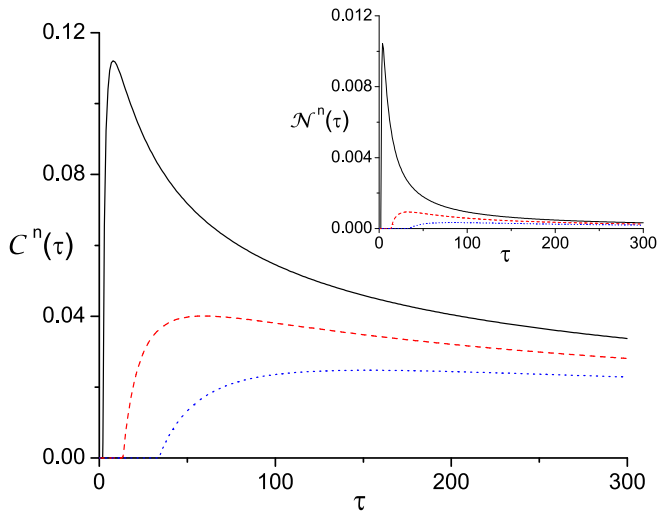}}
\caption{(Color online) Plots of ${\cal C}^n$ as a function of $\tau$ for 
$n=2$ (black solid line), $n=4$ (red dashed line) and $n=6$ (blue dotted line)
and for $\ga = \al=1$. The inset shows analogous plots for ${\cal N}^n$.} 
\label{fig1} \end{figure}

\section{Results}
\label{results1}

Eqs. (\ref{conc1}) and (\ref{nega1}) are the central results of this
work which lead to several conclusions about the two-spin
entanglement generated due to the quench for $XY$ model in a
transverse field. First, for odd $n$, $\la \si_i^z \si_j^z \ra = \la
\si_i^z \ra^2$ and $\la \si_i^+ \si_j^- \ra = 0$. Thus all
eigenvalues of $\rho^n (\si^y \otimes \si^y \rho^{n*} \si^y \otimes
\si^y)$ are equal to $\al_0 (1- \al_0)$ leading to ${\cal C}^n =0$.
All the eigenvalues of ${\bar \rho}$ are also positive; hence ${\cal
N}^n=0$. Thus {\it the quench generates entanglement only between
the even neighbor sites}. Second, for large $\tau_{\rm eff}$ and $n
\ll \sqrt{\tau_{\rm eff}}$, $\al_n$ scales similarly to the defect
density: $\al_n \sim \sqrt{\al/\tau}$. Using this, we find from Eqs.
(\ref{corr2}-\ref{corr3}) that $b_2^n \sim \sqrt{\al/\tau}$ and
$a_+^n a_-^n \sim \al/\tau$ which leads to ${\cal C}^n \sim
\sqrt{\al/\tau}$ and ${\cal N}^n \sim \al/\tau$. Thus, for slow
quenches, ${\cal C}^n$ (${\cal N}^n$) scales with the same (twice
the) universal exponent as the defect density [\onlinecite{antu,an1,ds3}].
This result relates two-spin entanglement generation for slow
quenches to defect production by such a process. Third, both the
concurrence and the negativity become non-zero for a {\it finite
critical quench rate $(\tau_c^{n})^{-1}$} (which is the solution of
$|b_2^n|^2 = a_+^n a_-^n$) above which there is no entanglement
between a site and its $n^{\rm th}$ neighbor. Solving this equation
numerically, we find $\ga^2 \tau_c^n =1.96,\,13.6\,{\rm and}\,33.8
$, for $n=2,\,4\,{\rm and}\,6$ when $\al=1$. We shall shortly see that 
for $\tau \le \tau_c^{2}$, the entanglement is entirely multipartite.

\begin{figure}
\rotatebox{0}{\includegraphics[width=3in]{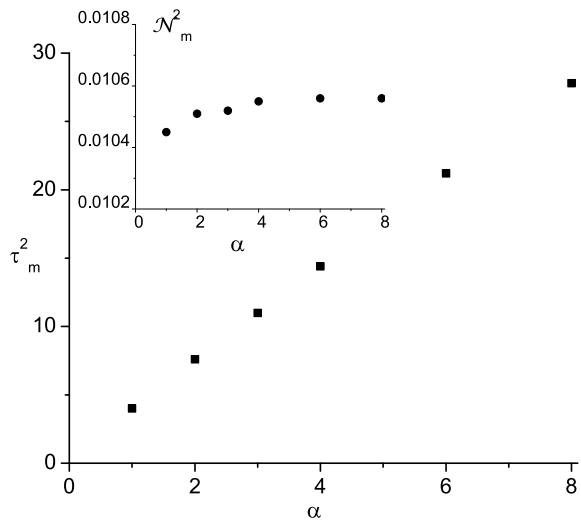}}
\caption{Variation of the position, $\tau_{m}^2$, of the maxima of
${\cal N}^2$ as a function of $\al$. The inset shows the values of the
maxima, ${\cal N}_m^2$, as a function of $\al$.} \label{fig2} \end{figure}

We now plot ${\cal C}^n$ and ${\cal N}^n$ as a function of $\tau$
for $n=2,4,6$ in Fig. \ref{fig1} for $\ga=\al=1$. From Fig.
\ref{fig1}, we find that ${\cal C}^n$ and ${\cal N}^n$ becomes
non-zero between $\tau= \tau_c^{n}$ and $\infty$. Further, the
ratios ${\cal C}^4/{\cal C}^2$ or ${\cal N}^4/{\cal N}^2$ {\it can
be selectively tuned between zero and 1} by tuning
$\tau$. The maximum values of both ${\cal C}^n$ and ${\cal N}^n$
decrease rapidly with $n$. For large $n \gg \sqrt{\tau}$, using the
properties of Toeplitz determinants used to compute the
spin-correlators in these systems, it can be shown that $\la \si_i^+
\si_{i+n}^-\ra \sim \exp(-n/\sqrt{\tau})$ [\onlinecite{lev1}]. Thus we
expect the entanglement to vanish exponentially for $n \gg
\sqrt{\tau}$. In Fig. \ref{fig2}, we plot the positions $\tau_m^{2}$
and the magnitudes ${\cal N}_m^2$ of the peaks in ${\cal N}^2$ as a
function of $\al$. We find that $\tau_m^{2}$ varies linearly with
$\al$ while ${\cal N}_m^2$ are independent of $\al$. This
behavior can be understood by noting that ${\cal C}^n$ and ${\cal
N}^n$ depend on $\al$ through $\tau_{\rm eff}$. The peak of
${\cal C}^n$ (${\cal N}^n$) occurs when $d{\cal C}^n/d\tau =
\al^{-1} d{\cal C}^n/d\tau_{\rm eff}=0$ ($d{\cal N}^n/d\tau =
\al^{-1} d{\cal N}^n/d\tau_{\rm eff}=0$). Thus the position of
the maxima, which is a solution to this equation, is given by
$\tau_{m\,\rm eff}^{n}= \al/\tau_m^n$ and hence varies linearly
with $\al$. The maximum value of the entanglement ${\cal
C}_m^{n}$ (${\cal N}_m^{n}$) is a function of $\tau_{m\,\rm
eff}^{n}$ alone; it does not change with $\al$.

\begin{figure} \rotatebox{0}{\includegraphics*[width=\linewidth]{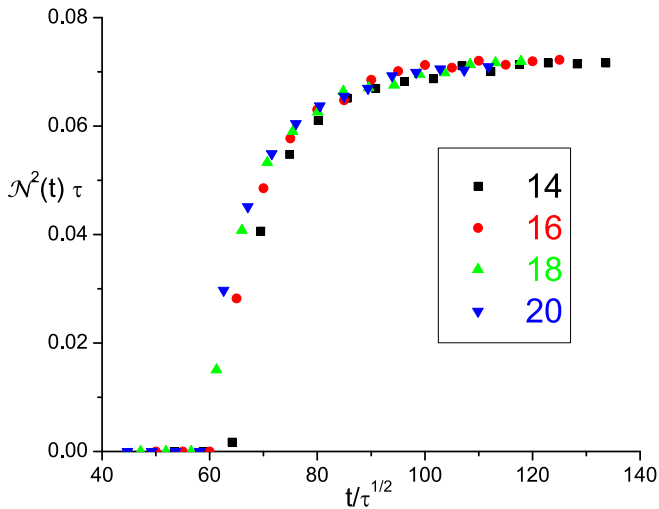}}
\caption{(Color online) Plot of scaled negativity ${\cal N}^2 (t)\tau$ as a 
function of $t/\sqrt{\tau}$ for $\al=\ga=1$ and several $\tau=14,16,18,20$, 
showing the time evolution of entanglement generated by the quench which 
started at $t_i=-250$.} \label{fig3} \end{figure}

Next, we study the time evolution of entanglement by
computing ${\cal N}^n$ at different times $t$ during the quench
for $\al=\ga=1$ and $n=2$. The plots of the scaled negativity
${\cal N}^2 (t)\tau$ as a function of the scaled time
$t/\sqrt{\tau}$ is shown in Fig. \ref{fig3} for several $\tau$.
From this plot, we find that for large enough $\tau$, the generation
of the entanglement starts when the system is near the QCP
and reaches an asymptotic value ${\cal N}^2$ as the quench takes
the system away from the quantum critical point. The collapse of all
the asymptotic values of the negativity to the same curve for all $\tau$
clearly confirms the scaling behavior of the entanglement discussed
earlier. Similar characteristics hold for ${\cal N}^{4,6}$.

\begin{figure}
\rotatebox{0}{\includegraphics*[width=3.4in]{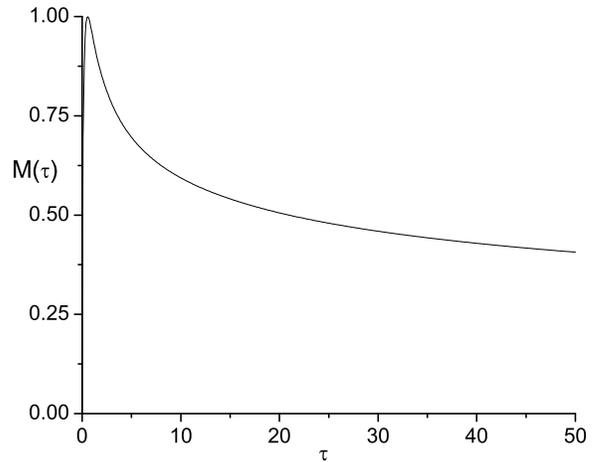}}
\caption{Plot of $M$ as a function of $\tau$ showing the evolution of the 
multipartite part of the entanglement with the quench time.} \label{fig4} 
\end{figure}

We now discuss the nature of the entanglement generated by the
quench. We note that the single-site density matrix of the spin
chain is $\rho_i = (I + \si^z \la \si_i^z\ra)/2$. Using Eq.
(\ref{corr1}), we find that the single-site concurrence ${\cal
C}^{(1)} = \sqrt{4 {\rm det} \rho_i} = 2 \sqrt{\al_0 (1-\al_0)}$
[\onlinecite{wootters2}] is finite for all $\tau$ except $\tau =0,\infty$
for which $\al_0=1,0$. Since ${\cal C}^{(1)}$ is a measure of the
entanglement of spin $i$ with all the other spins in the chain, we
conclude that the entanglement generated when $\tau \le \tau_c^{2}$
is entirely multipartite. The multipartite part of the entanglement
for any $\tau$ can be quantified as [\onlinecite{wootters2}]
\bea M &=& ({\cal C}^{(1)})^2 - \sum_{n\ge 2} ({\cal C}^n)^2. \eea
A plot of $M$ as a function of $\tau$ is shown in Fig.\ \ref{fig4},
where we have summed $C_n$ for $n \le 6$. We find that $M$ decreases
rather slowly; hence a determination of the fate of $M$ for large $\tau$
appears to be a rather difficult task. We note that for large $\tau$, $({\cal
C}^{(1)})^2 \to 0$ as $1/\sqrt{\tau}$, while $({\cal C}^n)^2 \to 0$
as $1/\tau$ for $\tau \gg n^2$. Hence $M$ may either go to $0$ as a
power of $\tau$ as $\tau \to \infty$, or may vanish at some large $\tau$ 
beyond which all the entanglement becomes bipartite. Differentiating 
between these two possibilities is left as a subject of future study.

An experimental verification of our theory would involve measuring the
two-spin correlation functions of an anisotropic $XY$ chain after
performing a quench. There are several compounds such as ${\rm
K_3Fe(CN)_6}$ ($J \simeq -0.23$K), ${\rm (NH_4)_2MnF_5}$ ($J \simeq
-12$K), and ${\rm Rb Fe Cl_3 \cdot 2H_20}$ ($J \simeq -35$K) where
the Ising limits of these chains are realized [\onlinecite{exp1}]. Similar
experiments, involving measurement of two-spin correlation functions
in equilibrium using neutron scattering, have recently been carried
out for square-lattice antiferromagnets, where short-range
entanglement between spins has been demonstrated [\onlinecite{chris1}].

\section{Conclusions}
\label{conclu1}

To conclude, we have shown that a controlled amount of entanglement
can be generated in an anisotropic $XY$ spin chain by performing a
power-law quench; such a process generates two-spin entanglement
only between sites which are even neighbors. The generated
entanglement is entirely multipartite when the quench is faster than
a critical rate; such states which have only multipartite
entanglement are quite uncommon. The entanglement between even
neighbors shows a scaling behavior for slow quenches, similar to the
scaling of defects. We note here that a similar calculation can be
performed for the two-dimensional Kitaev model [\onlinecite{kitaev1}] since
the exact two-spin correlations functions of the model during a quench
process with an arbitrary rate $\tau$ have been computed in Ref.
[\onlinecite{ds1}]. Such a calculation shows that the Kitaev model has zero 
bipartite entanglement (${\mathcal C}^{n}= {\mathcal N}^{n}=0$ for all $n$) 
for all quench times $\tau$, and the entire entanglement is always 
multipartite. This leaves us with the question of the relation between 
bipartite entanglement generated during a quench and the integrability and 
number of conserved quantities of the underlying system; this would be an 
interesting subject for future studies.

We thank S. Bandyopadhyay, I. Bose, and A. Sen(De) for helpful discussions.
D.S. thanks DST, India for financial support under Project No. 
SR/S2/CMP-27/2006.


\end{document}